\documentclass[preprint2]{aastex}
\shorttitle{Merger remnants in QSO host galaxies}
\shortauthors{Bennert et al.}

\begin{document}

\title{EVIDENCE FOR MERGER REMNANTS IN
EARLY-TYPE HOST GALAXIES OF LOW-REDSHIFT QSOS\altaffilmark{1}}

\author{Nicola Bennert\altaffilmark{2}, Gabriela Canalizo\altaffilmark{2,3}, 
Bruno Jungwiert\altaffilmark{2,4},
Alan Stockton\altaffilmark{5}, Fran\c{c}ois Schweizer\altaffilmark{6},
Chien Y. Peng\altaffilmark{7,8},
Mark Lacy\altaffilmark{9}}

\altaffiltext{1}{Based on observations made with the 
NASA/ESA {\em Hubble Space 
Telescope}, obtained at the Space Telescope Science Institute, which is 
operated by the Association of Universities for Research in Astronomy, Inc., 
under NASA contract NAS 5-26555. These observations are associated with 
program \# GO-10421.}

\altaffiltext{2}{Institute of Geophysics and 
Planetary Physics, University of California, Riverside, CA 92521;
nicola.bennert@ucr.edu, gabriela.canalizo@ucr.edu, bruno.jungwiert@ucr.edu}

\altaffiltext{3}{Department of Physics and Astronomy, University of California, Riverside, CA 92521}

\altaffiltext{4}{Astronomical Institute, Academy of Sciences of the Czech Republic,
Bo{\v c}n\'\i\ II 1401, 141 31 Prague 4, Czech Republic}

\altaffiltext{5}{Institute for Astronomy, University of Hawaii, 2680 Woodlawn Drive, 
Honolulu, HI 96822; stockton@ifa.hawaii.edu}

\altaffiltext{6}{Carnegie Observatories, 
813 Santa Barbara Street, Pasadena, CA 91101; schweizer@ociw.edu}

\altaffiltext{7}{Space Telescope Science Institute, 3700 San Martin Drive, 
Baltimore, MD 21218}

\altaffiltext{8}{NRC Herzberg Institute of Astrophysics, 5071 West Saanich
Road, Victoria, British Columbia, Canada V9E 2E7; cyp@nrc-cnrc.gc.ca}

\altaffiltext{9}{{\em Spitzer} Science Center, 
California Institute of Technology, Pasadena, CA 91125;
mlacy@ipac.caltech.edu}

\begin{abstract}
We present results from a pilot {\em HST} ACS deep 
imaging study in broad-band $V$ of 
five low-redshift QSO host galaxies classified in the literature as ellipticals.
The aim of our study is to determine whether these early-type hosts formed at 
high redshift and have since evolved passively, or whether
they have undergone relatively recent mergers that may be related to the triggering 
of the nuclear activity.
We perform two-dimensional modeling of the
light distributions to analyze the host galaxies' morphology.
We find that, while each host galaxy is reasonably 
well fitted by a de Vaucouleurs profile, the majority of them (4/5) reveal 
significant fine structure such as shells and tidal tails.  These structures 
contribute between $\sim$5\% and 10\% to the total $V$-band luminosity of each 
host galaxy within a region of $r \sim$ 3\,$r_{\rm eff}$ and are indicative
of merger events that occurred between a few hundred Myr and a Gyr ago.
These timescales are comparable to starburst ages in the 
QSO hosts previously inferred from Keck spectroscopy. Our results thus support
a consistent scenario in which most
of the QSO host galaxies suffered mergers with accompanying
starbursts that likely also triggered the QSO activity in some way,
but we are also left with considerable uncertainty on physical
mechanisms that might have delayed this triggering for several
hundred Myr after the merger.  
\end{abstract}


\section{INTRODUCTION}
Quasi-stellar objects (QSOs) are the most luminous active 
galactic nuclei (AGNs), believed to be powered by accreting supermassive black
holes (BH). The growing observational evidence that massive BHs exist not only 
in the centers of active galaxies, but also in inactive early-type galaxies
indicates that more than the mere presence of a massive
BH is needed to trigger the activity.
And while the ``bright quasar phase'' is now often cited as an essential 
phase in the evolution of galaxies \citep[e.g.,][]{spr05a},
we still do not have
a clear picture of what causes the triggering, or even what the 
necessary conditions are for the triggering to occur.

For more than two decades now, mergers have been suspected to be responsible
for triggering QSOs \citep{sto82,san88}.  There is an abundance of circumstantial
evidence in the literature connecting QSO activity at low redshifts with
mergers and interactions \citep[e.g.,][]{hec84,hut88,can01,urr07}.
Not only do a number of low-redshift QSOs show signs of interactions and mergers,
but there is also a tendency for QSO host galaxies 
to exist in only moderately rich environments 
\citep[e.g.,][see, however, \citealt{mar07} for a counterexample]{dre85,ell91,wol01}
consistent with the results of quasar clustering
studies, which suggests that quasars exist in haloes of mass $\sim$10$^{12}$ - 10$^{13}$
$M_{\odot}$ \citep{hop07a, daa07}.
In such environments, velocity dispersions are
moderate, and interactions  have their greatest effect.

At least in the case of QSOs found in ultraluminous infrared galaxies 
(ULIRGs), there is a close connection between mergers and QSO activity.
ULIRGs are essentially
always found in merging systems \citep{san96}, and far too many of them harbor QSOs
to be explained by the joint probability of finding a given $L^*$ or
brighter galaxy to be both a ULIRG and a QSO host galaxy by chance 
\citep{can01}.  So, at least for this subclass of QSO, some of 
the gas that the merger feeds into the central starburst also finds its way 
much deeper to the central black hole.

Nevertheless, recent high-resolution
imaging studies show that a large fraction of QSOs reside in hosts with 
relaxed morphologies
like elliptical galaxies \citep[e.g.,][]{dis95,bah97,mcl99,flo04}.  These 
observations have elicited claims that tidal interactions in QSO host 
galaxies are the exception rather than the rule, and that even in those
cases in which there are clear signs of interaction, the interactions are not
necessarily related to the nuclear activity 
\citep[e.g.,][hereafter D03]{dun03}.
From two-dimensional fitting of a well matched sample of
33 AGNs (radio-loud and radio-quiet QSOs as well as radio galaxies),
using images obtained with the Wide Field \& Planetary Camera 2 (WFPC2)
on board {\em HST}, D03
concluded that ``for nuclear luminosities $M_V < -23.5$, the hosts
of both radio-loud {\em and} radio-quiet AGN are virtually all massive
elliptical galaxies with basic properties that are indistinguishable
from those of quiescent, evolved, low-redshift ellipticals of comparable 
mass''. 

Such conclusions conflict with recent findings of 
significant amounts
of dust and cold molecular gas, as well as young stellar populations,
in many radio-loud and radio-quiet QSOs \citep[e.g.][]{mar99, can01,
eva01, sco03, tad05, bar06}. 
Although the hosts of most luminous AGNs are bulge-dominated,
they seem to be significantly bluer than inactive elliptical galaxies and
indeed show evidence for starbursts in the relatively recent
past (1--2 Gyr) \citep[e.g.,][]{kau03, san04, can06, sch06, jah07}.

In addition, both numerical simulations and observations show that merger remnants 
that have undergone violent relaxation have surface-brightness profiles that
follow $r^{1/4}$ laws \citep[e.g.,][]{too72,bar88,hib96}, even while still
showing clear signs of the tidal interactions \citep{rot04}.  
Thus, if QSO host galaxies are indeed relatively recent merger remnants,
they would most likely have elliptical profiles, but they would also show
fine structure indicative of past interactions.

Detecting fine structure in the already elusive QSO host galaxies 
is a challenging task.   Studies of nearby merger remnants
show that their fine structure is often faint compared to the galaxies
themselves.   For example, the total luminosity of the shells in
shell galaxies accounts only for 5--15\% of the total luminosity
of the galaxy \citep[e.g.,][]{for86,pri88,wil00},
and their detection often requires a resolution 
of at least $\sim$0.5 kpc \citep[e.g.,][]{sch92}. Therefore, 
detecting fine structure in QSO hosts requires high angular resolution  
and higher signal-to-noise (S/N) observations than those obtained in
previous studies.

Thus, we are conducting a study with deep (5 orbit) HST ACS and 
WFPC2 observations 
of a sample of QSO host galaxies that are classified as ellipticals.
We presented results for the first object in 
\citet[][hereafter Paper I]{can07}.
In the present paper, we describe results for the four remaining objects from 
our pilot ACS study of 5 objects. 
We will present results for the rest of the sample (14 additional objects) in 
subsequent papers.
Throughout this paper, we adopt $\Omega_{\lambda} = 0.7$, 
$\Omega_{m} = 0.3$ and $H_0$ = 71 km\,s$^{-1}$\,Mpc$^{-1}$.

\section{SAMPLE SELECTION}
Since the goal of our project is to investigate the possibility
that QSO host galaxies are formed through mergers even when
they do not show dramatic signs of tidal interactions,
we should ideally draw our sample from a sample of hosts
that have been classified as undisturbed elliptical galaxies.
The best such sample is that of D03.

From the detailed morphological study
of radio-loud and radio-quiet QSO host galaxies by D03,
we selected QSOs for which the host galaxies 
were classified as elliptical galaxies\footnote{Note that 
the D03 radio-loud and radio-quiet QSOs
were selected to be
statistically indistinguishable in their optical luminosity
and redshift.}.
Moreover, we selected those objects
for which existing deep Keck spectra 
did not reveal extended emission lines 
that could contaminate
our observations in the broad $V$-band filter (F606W).
We chose a sample
of five QSOs, three radio-quiet and two radio-loud.
The sample properties are summarized in Table~\ref{obstable}.

\section{OBSERVATIONS AND DATA REDUCTION}
All observations were obtained using 
the ACS Wide Field Channel (WFC) onboard the {\em HST}
with the broad $V$-band F606W filter ($\Delta \lambda$ = 2342\AA; 1 pixel
corresponds to 0.05\arcsec).
Five sets of dithered images were taken (with four subsets of 485--588 s integration
time each), yielding a total integration time of 10740--11464 s
per target (see Table~\ref{obstable}).

\begin{deluxetable}{lccccccc}
\tabletypesize{\scriptsize}
\tablecolumns{8}
\tablewidth{0pc}
\tablecaption{Details of Observation}
\tablehead{
\colhead{QSO} & \colhead{IAU designation} & \colhead{$\alpha$ (J2000)} & \colhead{$\delta$ (J2000)} & \colhead{$z$} & \colhead{Type}  & \colhead{Exp. Time} & \colhead{scale}\\
& & (hh mm ss) & ($\degr$ $\arcmin$ $\arcsec$) & & & (sec) & (kpc/$\arcsec$)\\
\colhead{(1)} & \colhead{(2)} & \colhead{(3)}  & \colhead{(4)} & \colhead{(5)}
& \colhead{(6)} & \colhead{(7)} & \colhead{(8)}}
\startdata
PHL\,909 & 0054+144 & 00 57 09.9 & +14 46 10 & 0.172 & RQQ & 10920 & 2.885\\
PKS\,0736+01 & 0736+017 & 07 39 18.0 & +01 37 05 & 0.191 &  RLQ & 11364 & 3.137\\
PG\,0923+201 & 0923+201 & 09 25 54.7 & +19 54 05 & 0.190 &  RQQ  & 11464 & 3.124\\
MC2\,1635+119 & 1635+119 & 16 37 46.5 & +11 49 49 & 0.146  & RQQ & 11432 & 2.520\\
OX\,169 & 2141+175 & 21 43 35.5 & +17 43 49 & 0.211 &  RLQ & 10740 & 3.392\\
\enddata
\tablecomments{
Col. (1): QSO. Col. (2): IAU designation. Col. (3,4): Optical positions
taken from NED. Col. (5): Heliocentric redshift as listed in NED. Col. (6): 
Classification as either radio-quiet QSO (RQQ) or radio-loud QSO (RLQ). 
Col. (7): Total integration time in sec. Col. (8): Physical scale in kpc/$\arcsec$.
}
\label{obstable}
\end{deluxetable}

In order to achieve background-limited images enabling
us to detect any existing faint structure in the QSO host galaxies,
we obtained individual frames with long integration times.
Consequently, the central pixels of the QSO nuclei are saturated
in these images. We discuss
the effects of the saturation in detail in the next section.

We re-calibrated the data manually,
correcting
the pipeline flat-fielded single exposures
for the bias-level offset between the
adjacent quadrants that is still present in
the final product of CALACS \citep{pav05}.
We used MultiDrizzle \citep{koe02} to combine
the individual images to the final
distortion and cosmic-ray corrected scientific image
(default values plus bits=8578 and a deltashift file with the offsets between
the images as determined from stars within the field-of-view (FOV)).
Note that no sky subtraction was performed during
MultiDrizzle.

\section{IMAGE ANALYSIS}
\label{process}
Different methods such as unsharp masking, creating
a structure map \citep{pog02}, and using the two-dimensional
galaxy fitting program GALFIT \citep{pen02}
were applied to look for any fine structure that might be present.
We here describe only the procedure
we adopted for the use of GALFIT as the other two methods did not
yield additional information 
(see Paper I for details on these methods).

For convolution with the point-spread function (PSF)
of the {\em HST} ACS optics, we created both an artificial
PSF star from 
TinyTim\footnote{Available at \\http://www.stsci.edu/software/tinytim/tinytim.html} 
(Version 6.3) at the same position
as each QSO and a PSF star from an observed star
on an ACS/WFC F606W image observed close
to the positions of the QSOs.
For the TinyTim PSF, 
we adopted a power law $F_{\nu} = \nu^{\alpha}$ with $\alpha$ = $-$0.3
which is the average value we obtain for the QSOs.\footnote{$(B-R)$ ranges between
0.1 and 0.7 as measured from nuclear Keck spectra.} 
The computed radius for this artificial PSF is 10\arcsec.

For the observed PSF star, we chose
a star with a signal-to-noise ratio (S/N) of 20,000 from the {\em HST}
archive. The star was observed
on 20 dithered images with a total exposure time of $\sim$8100\,s 
(GO-9433, data sets j6mf19* and j6mf21*) and at a position of the WFC
chip less than 5$\arcsec$ away from the position of each QSO;
note that the full ACS FOV is 3.4\arcmin~$\times$ 3.4\arcmin.
These images were processed in the same manner as described above
for the QSO observations. 
Finally, we eliminated a few faint objects surrounding the PSF star
and modified the PSF image to minimize any noise introduced 
during PSF subtraction and convolution procedures.
To adaptively smooth the PSF, 
we compared data values to the standard deviation $s$ of the surrounding
sky: (1) for data values $>$ 7$s$, we retained the unmodified PSF;
(2) for data values between 3$s$ and 7$s$, the image was smoothed with a Gaussian
kernel with $\sigma$ = 0.5 pixel; (3) for data values $<3s$, a Gaussian kernel 
of $\sigma$ = 2.0 pixel was used, and (4) for data values $<1s$ after this
last operation, the value was replaced with 0.
The radius of the observed PSF star is  $\sim$4\arcsec.
The full-width-at-half-maximum (FWHM) 
of the PSF is 0.1\arcsec~in both 
the artificial star and the observed PSF star, 
in close agreement with the FWHM of stars
on the ACS images close to the QSO.

The observed PSF star has $(B-R)$ $\simeq$ 1.4 mag, as inferred from USNO\footnote{United
States Naval Observatory} data.
To test chromatic effects
on the form of the PSF, we created a series of artificial PSF stars using 
TinyTim. We used power laws with $\alpha$ ranging from -4.0 to 0.9, 
corresponding to $(B-R)$ values between 2 mag and 0 mag (to account for the
observed range of $(B-R)$ of the QSOs and the star, plus possible
uncertainties). 
We also created PSFs using a blackbody spectrum
with temperatures between 3000 and 40000 K. 
We fitted these different PSFs 
and a host galaxy model to our QSOs. 
The derived host galaxy parameters vary by 2\%. 
We thus consider the chromatic effects
as negligible compared to the other uncertainties 
such as the PSF mismatch due to telescope 
or plate scale breathing, focus changes, 
PSF saturation (see below) etc. 

We subtracted the two different PSFs (star and TinyTim) from saturated
and unsaturated stars within the ACS FOV, for comparison.
The observed PSF star yielded significantly better results than the TinyTim PSF.
Thus, we used the observed PSF star for all the subsequent
fitting and, in the following, we simply refer to it as PSF or PSF star.
This exercise also allowed us to estimate the quality
of our PSF fit, taking into account the saturation of the QSO nucleus.
The central region within a radius of $\sim$1.7\arcsec~is 
strongly affected by the PSF subtraction. Any structure
within this region may be an artifact.

We created artificial images to test the influence of the 
PSF saturation on the derived QSO host galaxy parameters.
Since luminous quasars generally do not vary much over time
scales of a few years, we can compare our quasar luminosities
with D03 to estimate the degree of saturation in our images and
the potential systematic errors that might arise due to saturation.
This direct comparison shows that our AGNs may be saturated
by as much as 60\%. (Note that we define "60\% saturation"
in such a way that 60\% of the flux is lost due to saturation.)
In addition, we created model images where
the AGN has the same luminosity as in D03. We then saturated
the AGN by 60\%.  After masking out the central saturated pixels 
in the same manner as we analyzed
our data (see below), 
we find that our systematic errors can be as large as 
7\% for the magnitude, 30\% for the effective radius
of the different host galaxy components and up to 30\% for the
S{\'e}rsi{\'c} index. The PSF magnitude can be
underestimated by up to 90\% due to saturation. This explains the differences 
in AGN luminosity that we observe compared with D03 (see Table~\ref{results}).

GALFIT \citep{pen02} is a 2-dimensional fitting program based on $\chi$$^2$
minimization; it allows 
the user to simultaneously fit one or more objects in an image
with different model light distributions, such as S{\'e}rsi{\'c} \citep{ser68},
de Vaucouleurs \citep{dev48}, or exponential.
The following steps were performed.
We created a mask to exclude the saturated pixels in the center,
the diffraction spikes, surrounding objects,
and any visible fine structure in order to fit only
the smooth underlying host galaxy light distribution.
For the purpose of the $\chi$$^2$ minimization, pixels were weighted by
1/$\sigma$$^2$ with $\sigma$ based on Poisson statistics as estimated by
GALFIT. (Note that the weighting is however irrelevant 
for the central saturated pixel that were masked.)
A PSF and a S{\'e}rsi{\'c} function were fitted. 
We fitted the background sky and
bright close neighboring galaxies simultaneously, while faint neighboring objects
were simply masked.
This least-square fit was then subtracted
from the original image to obtain the residual image in which any
fine structure superposed on the smooth host galaxy light distribution is 
more readily visible.

To estimate the uncertainties caused by the sky determination,
we used two different approaches.
First, different sizes of the fitting region were used 
(up to the largest possible size)
with the sky varying freely.
Second, the sky was also determined independently
and held fixed to this value during alternative fits.

Note that a S{\'e}rsi{\'c} power law is defined as 
\begin{eqnarray*}
\Sigma (r) = \Sigma_{\rm eff} \exp \left[- \kappa_n \left(\left(\frac{r}{r_{\rm eff}}\right)^{1/n}-1\right)\right] \hspace{0.2cm} ,
\end{eqnarray*}
where $\Sigma_{\rm eff}$ is the pixel surface brightness at the effective radius $r_{\rm eff}$,
and $n$ is the S{\'e}rsi{\'c} index ($n=4$ for a de Vaucouleurs
profile, $n=1$ for an exponential
profile). 

To fit the host galaxies, we used three different models: 
(1) a single de Vaucouleurs profile
(S{\'e}rsi{\'c} with $n=4$), (2) a single S{\'e}rsi{\'c} profile ($n$ free)
and (3) a combination of a de Vaucouleurs plus exponential (S{\'e}rsi{\'c} with $n=1$) 
profile. 
If the sky value is kept fixed during the fitting process,
the fit for (1) has a total of nine free parameters: three parameters
for the fitting of the PSF ($\alpha$, $\delta$, magnitude),
and six free parameters for the fitting of
the de Vaucouleurs profile ($\alpha$, $\delta$, magnitude,
$r_{\rm eff}$, b/a, P.A.).
Model (2) has one additional free parameter for the index $n$.
Model (3) has six more free parameters than model (1).
These parameters correspond to the exponential
profile
($\alpha$, $\delta$, magnitude,
$r_{\rm eff}$, b/a, P.A.).
In all three cases, if the sky is also fitted, there are three additional
parameters (background and gradient in $\alpha$ and $\delta$).
In some cases, we used a S{\'e}rsi{\'c} profile to fit 
bright galaxies in close proximity to the QSO hosts (PHL\,909: six galaxies;
PG\,0923+201: seven galaxies, MC2\,1635+119: one galaxy).  
In those cases, the free parameters for these fits add to the number of
total free parameters.

\section{HOST GALAXY PROPERTIES AND FINE STRUCTURE}
In this section, we summarize the results
for all five QSOs in the pilot sample.
Although MC2\,1635+119 was discussed
in detail in Paper I, Figure~\ref{finalsbp}
shows its luminosity profile
which was not included in Paper I.
To facilitate comparisons, we also include the results for MC2\,1635+119 in the 
tables. 

\subsection{Host Galaxy Morphology}
\label{morphology}
\begin{figure}
\epsscale{1}
\plotone{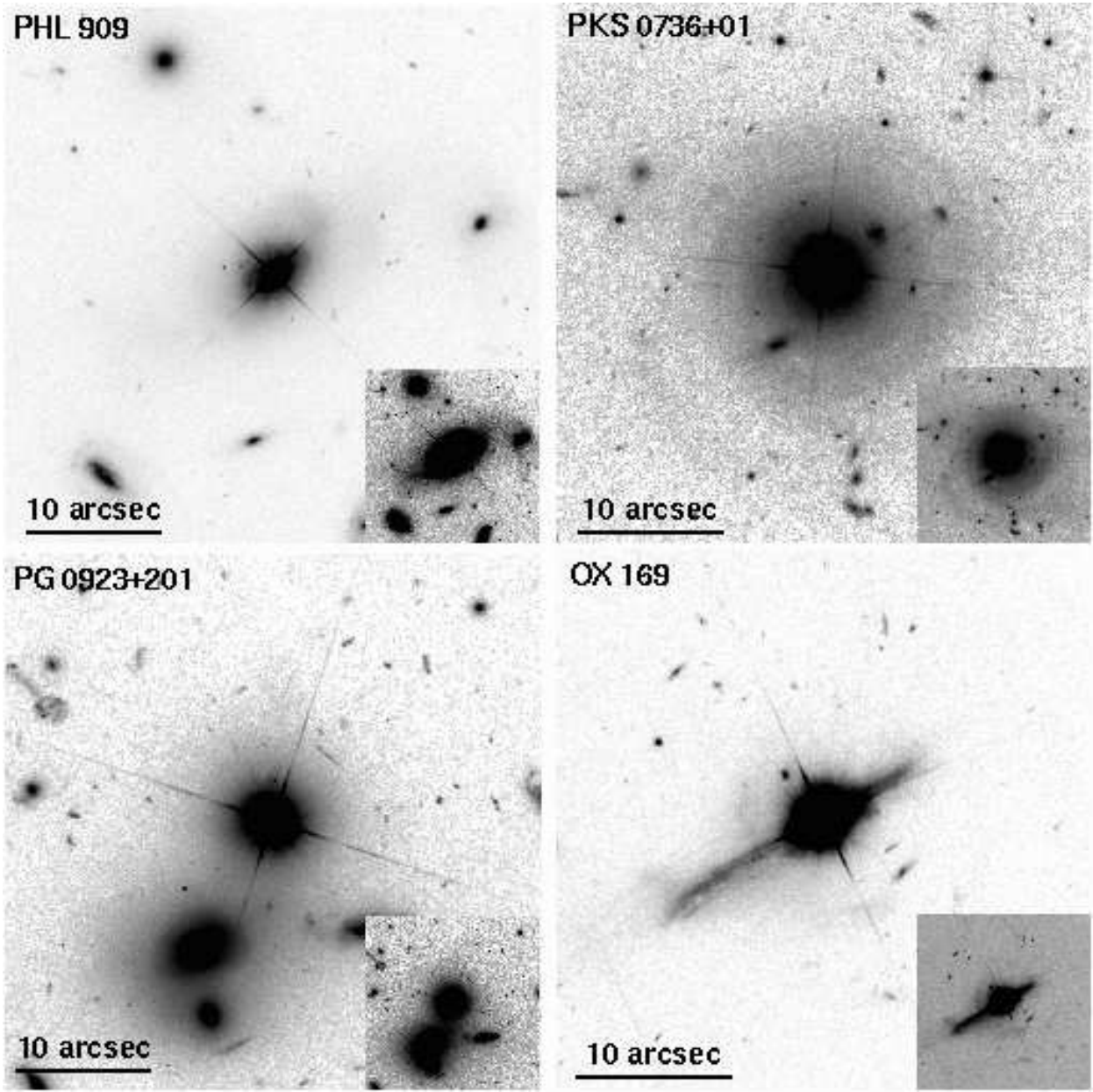}
\caption{ACS/WFC images of four QSO host galaxies, all smoothed with
a $\sigma = 0.5$ pixel Gaussian. 
In each image, north is up and east is to the left.
A harder stretch is shown as inset, Gaussian smoothed
with $\sigma = 1$ pixel (PHL\,909, OX\,169) and $\sigma = 2$ pixels
(PKS\,0736+01, PG\,0923+201), respectively.
Including MC2\,1635+119 (Paper I),
fine structure can be seen in at least four of the five
QSO host galaxies (PHL\,909: tidal tails and shells; PKS\,0736+201: spiral-like
structure;  OX\,169: disk-like structure, 
most likely tidal tail seen edge-on;
MC2\,1635+119: inner shells and outer arcs, see Paper I).
}\label{final}
\end{figure}

Figure~\ref{final} shows the ACS/WFC images for four of the five
QSOs in our sample: PHL\,909,
PKS\,0736+01, PG\,0923+201, and OX\,169.
Including MC2\,1635+119,
four of the five host galaxies reveal extended fine structure
on different scales and with different morphologies, ranging from
spectacular shells and arcs (MC2\,1635+119; Paper I), 
to tidal tails (PHL\,909) to
jet-like (OX\,169) and spiral-like (PKS\,0736+01) structure.
Only one object (PG\,0923+201) resides in a host galaxy without
any obvious fine structure.
We here describe the overall host-galaxy morphology
for each object in turn. Further details on each object, including
a comparison with results in the literature and
a description of neighboring objects, are given in Appendix~\ref{individual}.

The host galaxy of PHL\,909 shows 
a variety of fine structures at different
radii. A central ring- or disk-like structure 
($r \simeq$1.8\arcsec~-- 2.5\arcsec)
can be seen, possibly intersected by a dust lane
to the SE (Fig.~\ref{final}). 
To the NW, two shells are apparent. They occur at distances
of 2.5\arcsec~(P.A.~$\simeq$ 328) and 3\arcsec~(P.A.~$\simeq$ 298).
Diffuse outer material seems to form another ring 
or disk-like structure (Fig.~\ref{acs}, inset; $r \simeq$ 3.4\arcsec~-- 6\arcsec).
The most prominent fine structure consists of two tidal tails,
one extending from the
disk-like structure in the SE, curving towards
the N, and the other extending from the disk in the NW,
bending towards the south (Figs.~\ref{final} and~\ref{acs}). The western tidal
tail extends to a galaxy (``a''; Fig.~\ref{acs}) $\simeq$16\arcsec~to 
the NW (P.A.~$\simeq$ 284$\degr$). 
The tidal feature most likely corresponds to the faint
extended emission reported by \citet{mcl99}
which was also seen in the K-band image \citep{dun93}.
The companion (``a'') itself has two tidal tails, one reaching out in the direction
of the QSO and connecting with its tidal tail,
another tidal tail extending to the S up to 5.5\arcsec~(Fig.~\ref{final}).
Also the eastern tidal tail seems to be connected
to a very faint small companion (``d'') at 14\arcsec~distance
from the QSO (P.A.~$\simeq$ 98$\degr$), possibly a dwarf galaxy (Fig.~\ref{acs},
inset).

PKS\,0736+01 has been observed in the near-infrared 
K-band by \citet{dun93}.
The host is about 10\arcsec~in diameter
and is described as highly disturbed with a large area
of low surface-brightness nebulosity 
and several compact companions embedded.
In our deep {\em HST} ACS image, this nebulosity
is clearly visible and shows resolved structure that looks
like a faint face-on spiral. 
The spiral-like structure
extends out to a radius of 16\arcsec~($\simeq$ 50\,kpc),
i.e.~very large for a normal spiral galaxy.
Also, the pitch angle changes with radius, arguing
against a simple spiral-arm interpretation. 
Instead, what is observed here, may be
spatial wrapping of material from a
merger event (see discussion in Sec.~\ref{mergerage}).

PG\,0923+201 resides in a galaxy group of at least 6 galaxies,
some of which form closely interacting pairs, including the two
large elliptical galaxies to the south-east of the QSO 
(see Appendix~\ref{individual}).
However, the host galaxy itself does not reveal any 
obvious fine structure in the deep {\em HST} ACS image.

The host galaxy of MC2\,1635+119 has been described
in detail in Paper I. It reveals spectacular fine structure,
consisting of (at least) five interleaved inner 
shells (out to $r \simeq$ 13\,kpc) and 
several extended arcs out to a radius of $\sim$ 65\,kpc. 

The host galaxy of OX\,169 exhibits an extended linear structure
which has been reported by several authors  \citep{sto78,hut84,geh84,smi86,hec86}.
OX\,169 was studied by \citet{sto91} and \citet{hut92} who
described this jet-like structure in detail. \citet{sto91} showed
that it consists mainly of old stars at the same redshift.
This is supported by the prominence of the structure also
in the K-band image \citep{dun93}.
While the feature resembles a galactic disk seen edge-on, with a dust lane
in the SE, it is not centered on the QSO; almost 2/3 of it
lies on the SE side of the QSO.
Because of its strong asymmetry, \citet{sto91} favor an
interpretation as a tidal tail seen edge-on rather 
than a normal galactic disk. 
In our deep {\em HST} ACS image, the linear structure
is very prominent and extends over a projected distance of $\sim$68\,kpc.
The feature runs from NW to SE (P.A.~$\simeq$ 303$\degr$),
out to $\simeq$7.6\arcsec~in the NW and
$\simeq$12.5\arcsec~in the SE.
It is slightly curved and appears to continue across the nucleus.
A dust lane is clearly visible between 7.6\arcsec~and 11.5\arcsec~in the
SE side of the tail.
In addition to the prominent linear structure, the deep
{\em HST} ACS image reveals
faint extended structure for the first time: 
a shell-like feature in the NE (P.A.~$\simeq$ 83$\degr$)
at a distance of $\simeq$5\arcsec, plumes to the N of the SE end
of the tail (P.A.~$\simeq$ 110$\degr$) at a distance of $\simeq$11.5\arcsec~from the
center, and a fan extending from the SE tail towards
the SW. The latter extends out to 
a companion  (``a''; Fig.~\ref{acs})
to the SW (P.A.~$\simeq$ 235$\degr$)
at a distance of 6.1\arcsec~and may be a tidal stream.
The companion has an overall spiral-like
morphology which seems to be tidally stretched.
These findings provide strong support
for the interpretation of the linear structure
as a tidal tail seen edge on rather than it
being a normal galactic disk.
It is only natural to assume that all features have a common
origin during a merger event.

\onecolumn
\twocolumn

\subsection{Host Galaxy Properties}
As described in Section~\ref{process}, 
we modeled the QSO host galaxies with GALFIT
and fitted single de Vaucouleurs profiles,
single S{\'e}rsi{\'c} components,
and a combination of de Vaucouleurs and
exponential-disk profiles.
Varying the range of the sky parameter shows 
that the uncertainty in photometry can be up to 20\%,
the uncertainty in $r_{\rm eff}$ up to 10\%,
and the uncertainty in S{\'e}rsi{\'c} index up to 20\%.
These uncertainties add to the uncertainties from the PSF saturation
(see Sec.~\ref{process}).
The total uncertainties are $\sim$20\% for photometry,
$\sim$30\% in $r_{\rm eff}$ and $\sim$35\% in the S{\'e}rsi{\'c} index.
The results are summarized in Table~\ref{results}.

For all objects, good fits were achieved with all three
fitting functions mentioned above.
The fits improved slightly in going from a single de Vaucouleurs to
a single S{\'e}rsi{\'c} to a combination of de Vaucouleurs
plus exponential-disk profiles 
(see Table~\ref{results}). While this is expected simply 
because of the increasing number of free parameters,
the residual images do look smoother for the two-component
fits. 
In such a decomposition, the de Vaucouleurs
profile remains the dominant component, but
the exponential-disk component contributes between
6\% and 28\% of the total luminosity in $V$ of the hosts
(Table~\ref{results}).
The residual image did
improve significantly when adding a second component 
to the host in MC2\,1635+119 (Paper I).

Single-S{\'e}rsi{\'c} fits yielded a S{\'e}rsi{\'c} index 
around 4 for each of the host galaxies (ranging
between 2.9 and 6.5).
Our results for the surface-brightness profiles 
of the underlying host galaxies are thus in 
overall agreement with those of \citet{dun03}.
In Figure~\ref{finalall}, we show the residual images
derived by subtracting a GALFIT model consisting
of a single S{\'e}rsi{\'c} profile from the
observed images.

\subsection{Surface-Brightness Profiles}
We derived 
surface-brightness profiles for each object using the 
IRAF\footnote{IRAF (Image Reduction
and Analysis Facilities) is distributed by the National Optical Astronomy
Observatories, which are operated by AURA, Inc., under cooperative
agreement with the National Science Foundation.} task {\em ellipse}. 
The profiles were obtained for both
the observed surface brightness and the best-fitting
model using GALFIT and its subcomponents.
The resulting 1D profiles for the different fits
(single de Vaucouleurs, single S{\'e}rsi{\'c}, 
de Vaucouleurs plus exponential) are only marginally
different. Therefore, we only present the
best fitting model using a S{\'e}rsi{\'c} component in Figure~\ref{finalsbp}.

The deviation of the model from the data
within the central 1.7\arcsec~is most likely caused by the PSF subtraction.
This is especially pronounced
for PG\,0923+201, as this object has the brightest PSF
(see Table~\ref{results}), and the low S/N ratio in the outer
parts of the PSF star results in additional
noise in the data.
In general, data points at a higher
surface brightness than the model correspond to the
fine structure that was masked out during GALFIT fitting
but that was included for the creation of the 1D profiles.
This is clearly the case for OX\,169, for which
the tidal tail causes the deviation between data and
model beyond 5\arcsec.

\begin{figure}
\epsscale{1}
\plotone{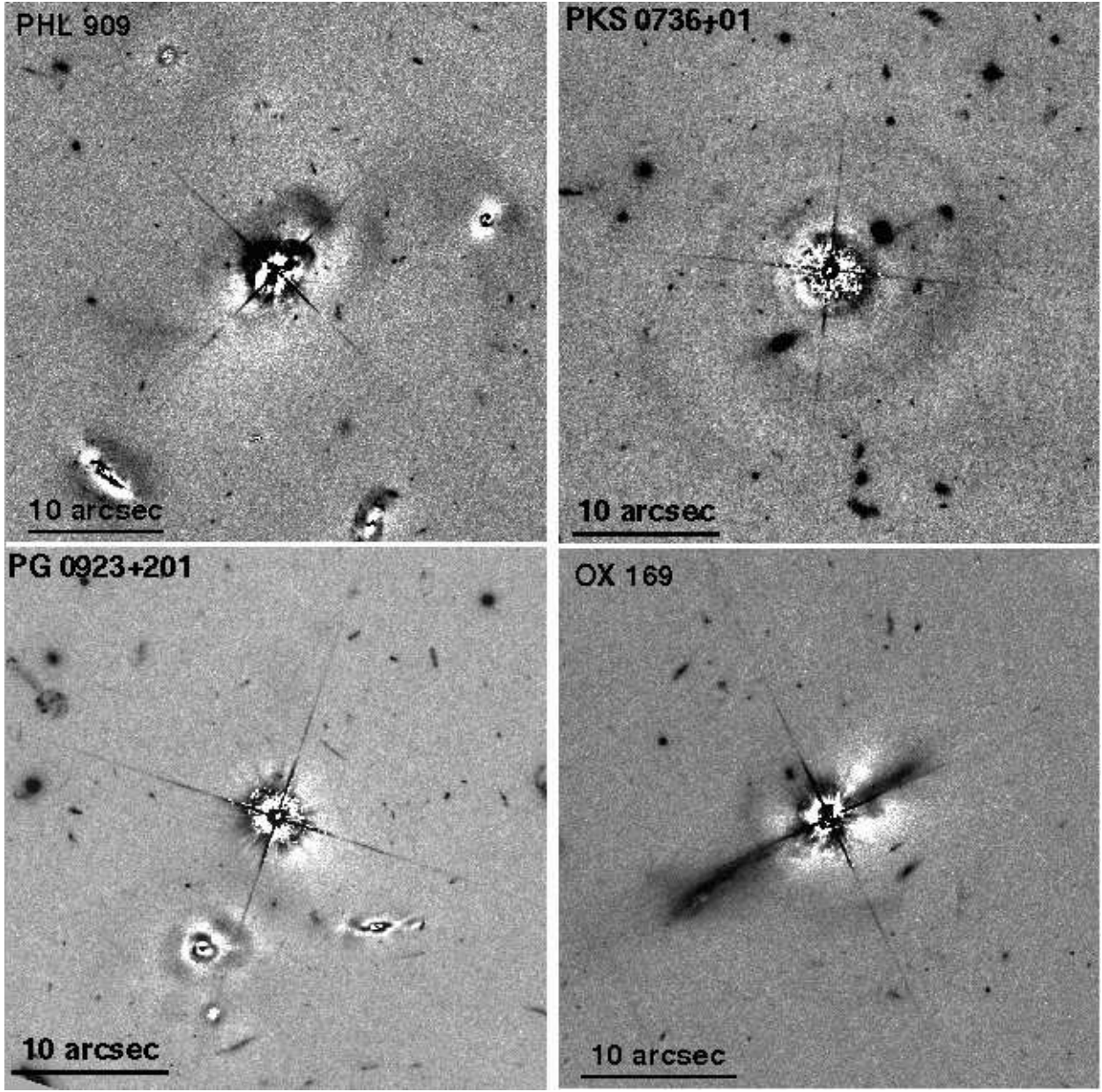}
\caption{Residual images derived
by subtracting a GALFIT model consisting of a single S{\'e}rsi{\'c}
profile.
In each image, north is up, east is to the left.
MC2\,1635+119 was presented in Paper I.
}\label{finalall}
\end{figure}

\begin{figure}
\epsscale{0.9}
\plotone{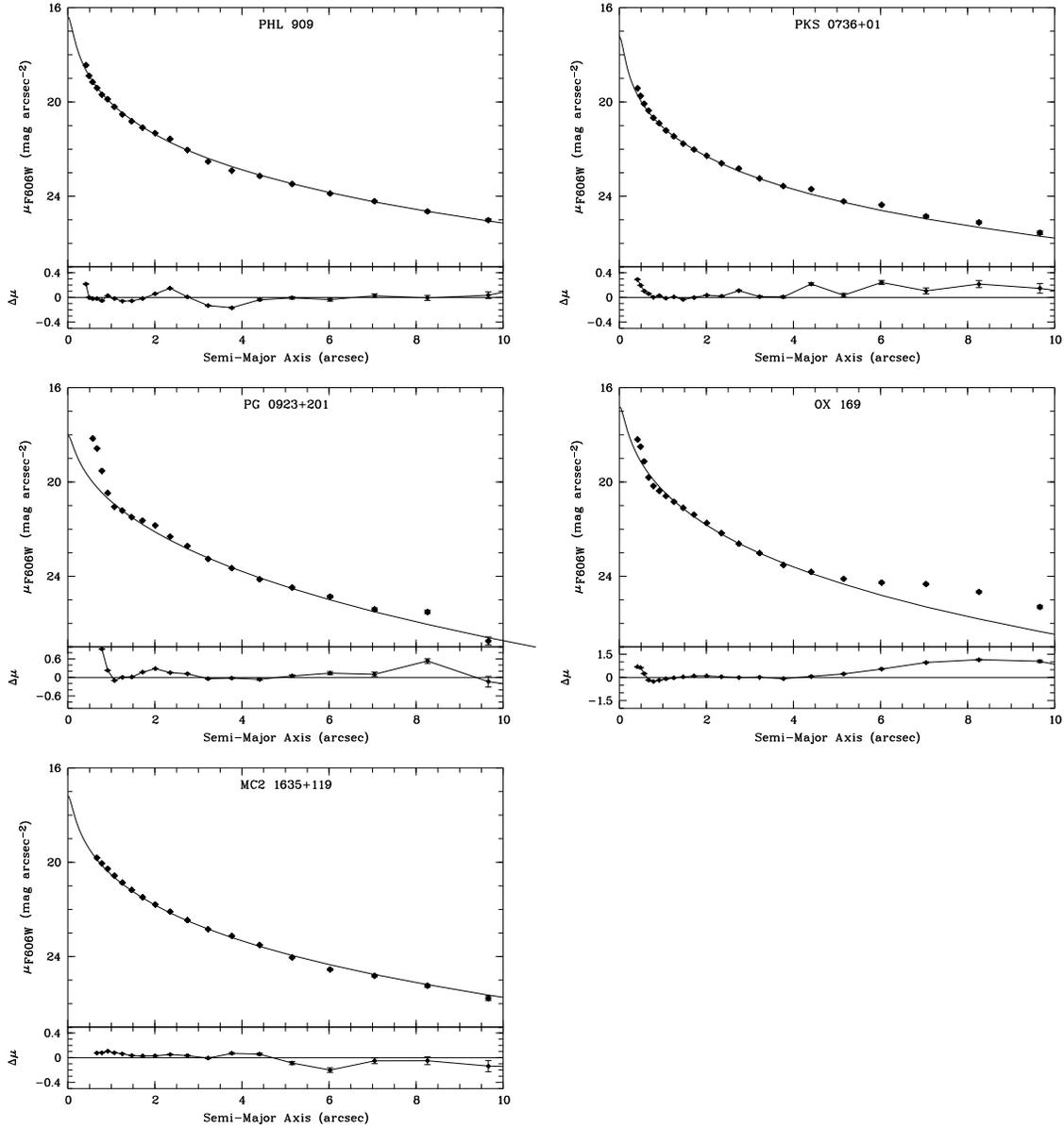}
\caption{Surface-brightness plots of observed and best-fitting model profiles
for all five QSO host galaxies.
The observed profile is shown as diamonds with error bars. 
The resulting profile of fitting a single S{\'e}rsi{\'c}
profile is shown as a solid line.
The residuals (fit $-$ data) are shown in the lower panels. 
}\label{finalsbp}
\end{figure}

The surface-brightness plot of the observed host profile
of PG\,0923+201
shows a bump peaking around 2\arcsec~that is not fitted by the model.
Since this feature lies just outside of the central region that is affected by
the PSF, we found it difficult to determine whether it is real or not.
However, it could reflect the presence of an inner disk.
If so, the outer radius of this disk would be approximately
8\,kpc.
The outer ``uplift'' of points over the best-fit profile
(around $r \sim 8$\arcsec) is most likely due to the companion
galaxy to the SSE and the relatively strong residuals that remain
after subtracting the GALFIT model for this galaxy (see also Fig.~\ref{finalall}).
The observed surface brightness for 
PHL\,909 also deviates slightly from the fit with a 
bump peaking around
2.3\arcsec.  This bump coincides with the 
ring-like structure seen in the image,
with an outer radius of $\sim$7\,kpc.

\begin{deluxetable}{lccccccccccc}
\tabletypesize{\scriptsize}
\tablecolumns{13}
\tablewidth{0pc}
\tablecaption{Results of Modeling the QSO Host Galaxies Using GALFIT}
\tablehead{
\colhead{QSO} & \colhead {Model} &
\colhead{Function} & \colhead{$\Delta$($\alpha$,$\delta$)} & \colhead{$m_{\rm F606W}$} 
& \colhead{$R$} & \colhead{$r_{\rm eff}$} & \colhead{$r_{\rm eff}$} & \colhead{S{\'e}rsi{\'c}} & \colhead{$b/a$} &
\colhead{P.A.} & \colhead{$\chi_\nu^2$} \\
\colhead{(fitted area)}& & & \colhead{($\arcsec$, $\arcsec$)} & \colhead{(mag)} & \colhead{(mag)} & \colhead{($\arcsec$)} & \colhead{(kpc)} & 
\colhead{index} & & \colhead{($\degr$)}\\
\colhead{(1)} & \colhead{(2)} & \colhead{(3)}  & \colhead{(4)} & \colhead{(5)}
& \colhead{(6)} & \colhead{(7)} & \colhead{(8)} & \colhead{(9)} & \colhead{(10)} & \colhead{(11)} & \colhead{(12)} }
\startdata
PHL\,909 & deV+Exp & PSF & (0,0) & 17.23 & \nodata & \nodata & \nodata & \nodata & \nodata & \nodata & 4.83  \\
(950$\times$950) & & deV & ($-$0.1,$-$0.1) & 17.08 & \nodata & 2.35 & 6.77 & 4 (fixed) & 0.57 & 130.4 & \nodata\\
& & Exp & ($-$0.96,1.45) & 19.88 & \nodata & 3.14 & 9.05 & 1 (fixed) & 0.57 & 111.0 & \nodata\\
& S & PSF & (0,0) & 18.18 & \nodata & \nodata & \nodata & \nodata & \nodata & \nodata & 5.14 \\
& & S & (0.1,$-$0.06) & 16.79 & \nodata & 2.53 & 7.30 & 5.6 & 0.63 & 130.2 & \nodata\\
& deV & PSF & (0,0) & 17.39 & \nodata & \nodata & \nodata & \nodata & \nodata & \nodata & 5.17  \\
& & deV & ($-$0.1,$-$0.05) & 16.96 & \nodata & 2.54 & 7.32 & 4 (fixed) & 0.61 & 130.4& \nodata\\
& deV (D03) & PSF & \nodata & \nodata & 15.5 & \nodata & \nodata & \nodata & \nodata& \nodata & \nodata \\
& & deV & \nodata & \nodata & 16.6 & 2.76 & 7.93 & 4 (fixed) & 0.61 & 131 & \nodata \\
& S (D03) & S & \nodata & \nodata & \nodata & \nodata & \nodata & 4 & \nodata & \nodata  & \nodata \\
\hline
\\*[-0.2cm]
PKS\,0736+01 & deV+Exp & PSF & (0,0) & 17.18 & \nodata & \nodata & \nodata & \nodata & \nodata & \nodata & 2.20 \\
(1051$\times$1051) & & deV & (0.01,$-$0.05) & 17.70 & \nodata & 1.63 & 5.10 & 4 (fixed) & 0.94 & 2.3& \nodata\\
& & Exp & ($-$0.94,$-$0.15) & 18.75 & \nodata & 7.09 & 22.23 & 1 (fixed) & 0.80 & 86.2& \nodata\\
& S & PSF & (0,0) & 16.07 & \nodata & \nodata & \nodata & \nodata & \nodata & \nodata & 2.25 \\
& & S & ($-$0.04,0) & 17.08 & \nodata & 4.27 & 13.41 & 6.5 & 0.98 & 32.7 & \nodata\\
& deV & PSF & (0,0) & 16.86 & \nodata & \nodata & \nodata & \nodata & \nodata & \nodata & 2.27 \\
& & deV & ($-$0.04,0) & 17.35 & \nodata & 3.30 & 10.34 & 4 (fixed) & 0.98 & 39.2& \nodata\\
& deV (D03) & PSF & \nodata & \nodata & 16.2 & \nodata & \nodata & \nodata & \nodata& \nodata & \nodata\\
& & deV & \nodata & \nodata & 16.9 & 3.27 & 10.21 & 4 (fixed) & 0.97 & 26 & \nodata\\
& S (D03) & S & \nodata & \nodata & \nodata & \nodata & \nodata & 5.3 & \nodata & \nodata  & \nodata \\
\hline
\\*[-0.2cm]
PG\,0923+201 & deV+Exp & PSF & (0,0) & 16.39 & \nodata & \nodata & \nodata & \nodata & \nodata & \nodata & 2.59 \\
(1051$\times$1051) & & deV & ($-$0.04,$-$0.02) & 17.40 & \nodata & 1.44 & 4.49 & 4 (fixed) & 0.96 & 174.2  & \nodata\\
& & Exp & (1.02,0.75)   & 20.45  & \nodata & 2.26 & 7.07 & 1 (fixed) & 0.51 & 146.1 & \nodata \\
& S & PSF & (0,0) & 16.15 & \nodata & \nodata & \nodata & \nodata & \nodata & \nodata & 2.76 \\
& & S & (0.07,$-$0.03) & 17.48 & \nodata & 1.81 & 5.66 & 2.9 & 0.97 & 42.1 & \nodata\\
& deV & PSF & (0,0) & 16.36 & \nodata & \nodata & \nodata & \nodata & \nodata  &  \nodata & 2.78 \\
& & deV & ($-$0.06,0.01) & 17.31 & \nodata & 1.68 & 5.2 & 4 (fixed) & 0.97 & 42.9 & \nodata\\
& deV (D03) & PSF & \nodata & \nodata & 15.7 & \nodata & \nodata & \nodata & \nodata  & \nodata & \nodata\\
& & deV & \nodata & \nodata & 17.2 & 2.02 & 6.30 & 4 (fixed) & 0.98 & 38 & \nodata\\
& S (D03) & S & \nodata & \nodata & \nodata & \nodata & \nodata & 3.3 & \nodata & \nodata  & \nodata \\
\hline
\\*[-0.2cm]
MC2\,1635+119 & deV+Exp & PSF & (0,0) & 17.58 & \nodata & \nodata & \nodata & \nodata & \nodata & \nodata & 3.49 \\
(481$\times$481) & & deV & ($-$0.03,0.04) & 17.50 & \nodata & 1.12 & 2.82 & 4 (fixed) & 0.71 & 57.9& \nodata\\
& & Exp & (0.53,0.36) & 18.81 & \nodata & 5.48 & 13.8 & 1 (fixed) & 0.82 & 27.7& \nodata\\
& S & PSF & (0,0) & 17.02 & \nodata & \nodata & \nodata & \nodata & \nodata & \nodata & 3.91 \\
& & S & ($-$0.03,0.04) & 17.18 & \nodata & 2.47 & 6.23 & 4.95 & 0.71 & 55.9 & \nodata\\
& deV & PSF & (0,0) & 17.17 & \nodata & \nodata & \nodata & \nodata & \nodata &  \nodata & 3.93\\
& & deV & ($-$0.03,0.04) & 17.27 & \nodata & 2.05 & 5.16 & 4 (fixed) & 0.71 & 56.2& \nodata\\
& deV (D03) & PSF & \nodata & \nodata & 18.1 & \nodata & \nodata & \nodata & \nodata& \nodata & \nodata\\
& & deV & \nodata & \nodata & 16.8 & 2.28 & 5.73 & 4 (fixed) & 0.69 & 56& \nodata\\
& S (D03) & S & \nodata & \nodata & \nodata & \nodata & \nodata & 5.6 & \nodata & \nodata  & \nodata \\
\enddata
\tablecomments{Col. (1): QSO and, in parentheses, the size of the 
fitted area in pixels. Col. (2) GALFIT model (deV = de Vaucouleurs, Exp = Exponential, S = S{\'e}rsi{\'c}).
Col. (3): Individual components used.
Col. (4): Offsets with respect to the PSF. Col. (5): Integrated
apparent magnitude in the F606W filter. Note that the magnitudes $m_{\rm F606W}$
given for the PSF underestimate the AGN component as the PSF was saturated
in all cases; see text for details.
Col. (6): Integrated apparent
magnitude converted from F675W to Cousins $R$ band as given by D03.
Col. (7,8): Effective radius in \arcsec~and kpc, respectively.
Note that although we concluded in Section~\ref{process} that
the central region within a radius of $\sim$1.7\arcsec~is 
still strongly affected by the PSF subtraction, all $r_{\rm eff}$ given
in this table are more than 10 times the full-width-at-half-maximum (FWHM)
of the ACS PSF.
Col. (9): S{\'e}rsi{\'c} index. Col. (10): Axis ratio.
Col. (11): Position angle (east of north). Col. (12): $\chi^2$ 
per degree of freedom, $\nu$. $\chi^2_{\nu}$
corresponds to the model in column 2.\\
Results from D03 are listed for
comparison. The $r_{\rm eff}$ given here for the D03 results 
were recalculated according to the worldmodel
adopted in this paper. The P.A. given here for the D03 results were derived
by adding the P.A. given in their Table 3 to the orientation of the spacecraft
(as the P.A. given in their Table 3 is apparently not corrected for it).
}
\label{results}
\end{deluxetable}

\begin{deluxetable}{lccccccccccc}
\tabletypesize{\scriptsize}
\tablenum{2 - continued}
\tablecolumns{13}
\tablewidth{0pc}
\tablecaption{}
\tablehead{
\colhead{QSO} & \colhead {Model} &
\colhead{Function} & \colhead{$\Delta$($\alpha$,$\delta$)} & \colhead{$m_{\rm F606W}$} 
& \colhead{$R$} & \colhead{$r_{\rm eff}$} & \colhead{$r_{\rm eff}$} & \colhead{S{\'e}rsi{\'c}} & \colhead{$b/a$} &
\colhead{P.A.} & \colhead{$\chi_\nu^2$} \\
\colhead{(fitted area)}& & & \colhead{($\arcsec$, $\arcsec$)} & \colhead{(mag)} & \colhead{(mag)} & \colhead{($\arcsec$)} & \colhead{(kpc)} & 
\colhead{index} & & \colhead{($\degr$)}\\
\colhead{(1)} & \colhead{(2)} & \colhead{(3)}  & \colhead{(4)} & \colhead{(5)}
& \colhead{(6)} & \colhead{(7)} & \colhead{(8)} & \colhead{(9)} & \colhead{(10)} & \colhead{(11)} & \colhead{(12)} }
\startdata
OX\,169 & deV+Exp & PSF & (0,0) & 17.80 & \nodata & \nodata & \nodata & \nodata & \nodata & \nodata & 3.45\\
(900$\times$900) & & deV & (0,0) & 17.68 & \nodata & 1.82 & 6.17 & 4 (fixed) & 0.47 & 120.4& \nodata\\
& & Exp & (0,0.15) & 19.34 & \nodata & 1.23 & 4.17 & 1 (fixed) & 0.66 & 18.5& \nodata \\
& S & PSF & (0,0) & 17.54 & \nodata & \nodata & \nodata & \nodata & \nodata & \nodata & 3.66\\
& & S & (0,0) & 17.42 & \nodata & 1.30 & 4.41 & 3.9 & 0.64 & 122.9 & \nodata\\
& deV & PSF & (0,0) & 17.56 & \nodata & \nodata & \nodata & \nodata & \nodata & \nodata & 3.66 \\
& & deV & (0,0) & 17.41 & \nodata & 1.29 & 4.36 & 4 (fixed) & 0.64 & 123.0 & \nodata \\
& deV (D03) & PSF & \nodata & \nodata & 15.9 & \nodata & \nodata & \nodata & \nodata& \nodata & \nodata\\
& & deV & \nodata & \nodata & 17.3 & 1.88 & 6.35 & 4 (fixed) & 0.47 & 121& \nodata \\
& S (D03) & S & \nodata & \nodata & \nodata & \nodata & \nodata & 3.6 & \nodata & \nodata  & \nodata \\
\enddata
\tablecomments{Col. (1): QSO and, in parentheses, the size of the 
fitted area in pixels. Col. (2) GALFIT model (deV = de Vaucouleurs, Exp = Exponential, S = S{\'e}rsi{\'c}).
Col. (3): Individual components used.
Col. (4): Offsets with respect to the PSF. Col. (5): Integrated
apparent magnitude in the F606W filter. Note that the magnitudes $m_{\rm F606W}$
given for the PSF underestimate the AGN component as the PSF was saturated
in all cases; see text for details.
Col. (6): Integrated apparent
magnitude converted from F675W to Cousins $R$ band as given by D03.
Col. (7,8): Effective radius in \arcsec~and kpc, respectively.
Note that although we concluded in Section~\ref{process} that
the central region within a radius of $\sim$1.7\arcsec~is 
still strongly affected by the PSF subtraction, all $r_{\rm eff}$ given
in this table are more than 10 times the full-width-at-half-maximum (FWHM)
of the ACS PSF.
Col. (9): S{\'e}rsi{\'c} index. Col. (10): Axis ratio.
Col. (11): Position angle (east of north). Col. (12): $\chi^2$ 
per degree of freedom, $\nu$. $\chi^2_{\nu}$
corresponds to the model in column 2.\\
Results from D03 are listed for
comparison. The $r_{\rm eff}$ given here for the D03 results 
were recalculated according to the worldmodel
adopted in this paper. The P.A. given here for the D03 results were derived
by adding the P.A. given in their Table 3 to the orientation of the spacecraft
(as the P.A. given in their Table 3 is apparently not corrected for it).
}
\end{deluxetable}

\begin{deluxetable}{lccc}
\tabletypesize{\small}
\tablecolumns{4}
\tablewidth{0pc}
\tablecaption{Contribution of Fine Structure to Host Galaxy Light}
\tablehead{
\colhead{QSO} & \colhead{fs (3\,$r_{\rm eff}$)}  & \colhead{$\Sigma_{\rm fs}$ (3\,$r_{\rm eff}$)} & \colhead{$\Sigma_{\rm fs}$ (max)}\\
& \colhead{$\%$}  & \colhead{mag\,\,arcsec$^{-2}$} & \colhead{mag\,\,arcsec$^{-2}$}\\
\colhead{(1)} & \colhead{(2)} & \colhead{(3)}  & \colhead{(4)}}
\startdata
PHL\,909 &  7.8 & 29.2 & 24.8\\
PKS\,0736+01 &  8.2 & 29.6 & 26.5 \\
MC2\,1635+119 &  6.3 & 28.3 & 25.2 \\
OX\,169 &  13.3 & 29.5 & 23.7 \\
\enddata
\tablecomments{
Col (1): QSO. Col. (2): Percentage of flux contained
in the observed fine structure within 3\,$r_{\rm eff}$.
Col. (3): Surface-brightness magnitude
of the fine structure in the F606W filter within 3\,$r_{\rm eff}$.
Col. (4): Average surface-brightness magnitude within an area of 
2\arcsec$\times$2\arcsec~around the maximum
of the fine structure.
}
\label{percentage}
\end{deluxetable}

\onecolumn
\twocolumn

\onecolumn
\twocolumn

\subsection{Fine Structure}
To estimate the luminosity within the various types of
fine structure compared to the total luminosity
of the host galaxy, we
created a mask that included all the light within 
an annulus with an inner radius 1.7\arcsec~and an outer radius
3\,$r_{\rm eff}$ (as taken from a single de Vaucouleurs
fit, see Table~\ref{results}).
The image was multiplied by
this mask (good=1, bad=0), and the total counts in
the product were summed.  This was done both for the GALFIT residual image
($f_{\rm fine structure}$) obtained by subtracting the GALFIT model
of a de Vaucouleurs + exponential profile, and for the GALFIT model itself
($f_{\rm galaxy}$). We then computed the ratio $f_{\rm fine structure}$/$f_{\rm galaxy}$.
The fractional luminosity of the extended fine structure in such an annulus
ranges between 2\% and 5\%.
To account for regions that were over-subtracted by GALFIT,
we also used a slightly different approach:
Applying the same general method as above, we only summed
the pixels that are $> -1 \sigma$ of the sky background.
Then, the fractional luminosity of the extended fine structure
ranges between 6\% and 13\%.
The results are summarized in Table~\ref{percentage}.

However, note that these values give only the average fraction within 3\,$r_{\rm eff}$.
Locally, the surface-brightness magnitude of the fine structure relative to
the underlying model can be much higher.
The surface-brightness magnitude within an area of 2\arcsec$\times$2\arcsec~centered
on the maximum of the fine structure lies
between $\sim$24 and $\sim$26 mag\,arcsec$^{-2}$ (as derived from the GALFIT residual image)
for the four objects with fine structure (Table~\ref{percentage}).

\section{DISCUSSION}
\label{mergerage}
We find significant fine structure in the deep {\em HST} ACS images 
of at least four of the five QSO host galaxies 
(PHL\,909, PKS\,0736+01, MC2\,1635+119, and OX\,169) in our pilot sample.
While the prominent linear structure in OX\,169 has been previously reported by several 
authors \citep{sto78,hut84,geh84,smi86,hec86,mcl99}, the structure in the other
three objects is revealed here for the first time.
Only one of the objects studied, PG\,0923+201, does not show
any obvious fine structure.

Our findings are supported by the 
WFPC2 images of D03 and \citet{mcl99} 
since some of the structure that we find is also visible in
their images:
Part of the north-western tidal tail in PHL\,909 is visible (Fig.\ A11,
\citealt{mcl99});
some of the spiral-like structure in PKS\,0736+01 can be seen
(Fig.\ A6 in \citealt{mcl99}); and the shell in MC\,1635+119 labeled
``d'' in Fig.~3 of Paper I 
is visible to the north-east of the QSO nucleus
at a distance of $\sim$4\arcsec~(Fig.\ A18 in \citealt{mcl99}).
However, the depth reached in our ACS images
is needed to positively identify the ``fuzz'' seen in these
earlier images as real.
Our results clearly show the need of high S/N imaging for an
accurate interpretation
of QSO host galaxy morphologies.

The host galaxy of PG\,0923+201 does not show obvious signs
of fine structure. However, the surface-brightness profile reveals
a bump in the inner 2\arcsec~that could reflect the presence
of an inner disk. Interestingly, PG\,0923+201 resides in a galaxy group of
at least 6 galaxies. It is the only QSO in our sample that has
been observed with the {\em Spitzer} Space Telescope, both
with the Infrared Spectrograph (IRS) and the
Multiband Imaging Photometer for Spitzer (MIPS; 24$\mu$m).
Neither FIR emission nor Polycyclic-Aromatic Hydrocarbon (PAH) features 
have been detected \citep{shi07}.
Interestingly, this lack of signs of star-formation activity in the infrared
goes hand-in-hand with the lack
of obvious signs for recent merger activity.

Much of the fine structure observed in our deep {\em HST} ACS
images of the other four QSO host galaxies is indicative of merger events in the
relatively recent past. 
The possible type and timescales for the merger
in MC2\,1635+119 were discussed in detail in Paper I.
Briefly, we found that the fine structure seen in the host galaxy of this object
was most likely produced by a strong tidal interaction within the last 
$\sim$1\,Gyr.   On the one hand,  the regularity of the inner shell structure 
is suggestive  of a near radial collision of a dwarf galaxy
with a giant elliptical galaxy in a minor merger.
On the other hand, the arcs observed at much larger distances than 
the inner shells, and other tidal debris off-axis from the direction of 
the encounter implied by the inner shells, are more indicative of a major
merger.   Using simple $N$-body simulations, we estimated that the time scale
needed to form the observed structure 
is less than $\sim$1.7 Gyr (see Paper I for details).
In comparison, the spectrum of the host galaxy is dominated
by a population of intermediate-age ($\sim$1.4 Gyr) stars, indicative of
a strong starburst that may have occurred during the merger event.
Thus, the observed QSO activity 
may have been triggered in the recent past either by a minor merger
or by debris from an older ($\sim$Gyr) major merger that is currently 
``raining'' back into the central regions of the merger remnant.

The prominent extended structure observed in OX\,169 is most likely
a tidal tail seen edge on.
Prominent tidal tails are also observed in the host of PHL\,909.
The tails in both QSO hosts 
are strongly suggestive of relatively recent major merger events,
where at least some of the parent galaxies were disk galaxies.

The structure surrounding PKS\,0736+201 resembles that of the
Seyfert 1.5 galaxy NGC\,5548, an Sa galaxy with ripples, tidal arms and a faint
long tidal tail that has been described in detail
by \citet{sch88}. The observed structure in NGC\,5548 has been interpreted
in terms of the spatial wrapping of material from an obliquely infalling
companion. When seen in projection, such material can appear in the form of
relatively sharp-edged ripples \citep{sch88}.
Compared with NGC\,5548, the faint spiral-like structure
seen in PKS\,0736+01 is more extended ($\sim$ 50\,kpc vs. 10\,kpc)
and more regular.

Structure similar to that
observed in PKS 0736+201 can be reproduced
in numerical simulations of minor mergers.
For example, the merger of a dwarf elliptical galaxy
with a large spiral and a total mass ratio of 1:8
results in very similar structure $\sim$1\,Gyr after the
first passage of the two galaxies, but also again
at later stages in the merger event (\citealt{you07};
T. J. Cox, private communication).

By comparing the structure observed in these hosts to
results from numerous published $N$-body simulations of mergers
(e.g., \citealt{hib95}), we know that the ages of these
tidal features are likely of the order of a few orbital times.
Given the masses of giant elliptical galaxies and
the distances at which we observe these features,
this time span ranges from a few hundred Myr to about 1 Gyr.
In the cases in which we can clearly see tidal tails, we can infer
that the dynamical timescales cannot be much less than $\sim 250$ Myr,
since that is roughly the amount of time that it would take for stars with
typical orbital velocities ($\sim 200$\,km\,s$^{-1}$) to travel in a straight
line to the projected maximum distances at which we observe the features
(e.g., $\sim$ 50 kpc for PHL\,909 or $\sim$ 42 kpc for OX\,169).
On timescales longer than $\sim$1 Gyr, tidal tails are expected to rapidly
disperse and lose contrast.

Further evidence that these are relatively evolved mergers is the fact
that enough time has elapsed since the initial tidal encounter
for each host to acquire an essentially elliptical
morphology. While it is possible that some of the objects,
such as MC\,1635+119 and PG\,0923+201, may have started off as
elliptical galaxies, at least OX\,169 and PHL\,909 must have
resulted from the merger of disk galaxies.   The fact that
they are at the present time reasonably well fitted by de Vaucouleurs
profiles implies merger timescales on the order of a Gyr.

To address the question of whether the AGN activity may
indeed have been triggered by the merger events, we need to obtain
estimates of the different timescales involved.
How do the ages of the stellar populations compare to the time elapsed since
the merger events, and what does that imply for the QSO duty cycle?
What is the connection between
merger, star formation, and BH accretion?

As mentioned above, the timescales for the merger events observed in
our sample is of the order of a Gyr or less.   G. Canalizo \& A. Stockton 
(2008, in preparation) find similar timescales for strong starburst events in
the host galaxies of these QSOs.  By modeling deep Keck spectra of these
objects, they find traces of major starburst episodes in intermediate-age
starburst populations that involve a substantial fraction of the stellar mass.
These starburst ages range from $\sim0.7$ Gyr (OX\,169) to 
2.2 Gyr (PKS\,0736+017).  Hence, it seems possible that the starbursts were
triggered by the merger events.

The most difficult timescale to estimate is that of the duration
of the AGN activity.
Current estimates of QSO lifetimes range between 10$^6$ and
10$^8$ years and are derived primarily from demographics
(see review by \citealt{mar04}).
For example, from theoretical calculations \citet{yu02}
estimate a mean lifetime
for a luminous QSO ($L_{\rm bol}$ $\ge$ 10$^{46}$\,erg\,s$^{-1}$)
of (3--13) $\times$ 10$^7$ yr which is comparable
to the Salpeter time scale ($\sim$ 4.5 $\times$ 10$^7$ yr
for $\epsilon$ = 0.1 and $L/L_{\rm edd}$ = 1; \citealt{sal64,mar04}).
Apart from demographics, there have been several
attempts to deduce the AGN duty cycle for individual galaxies.
For example, the length of radio jets 
can yield a lower estimate of QSO lifetimes,
depending on the expansion speed of the jet
\citep[e.g.,][]{wil78,sch95,blu99}. However, this method cannot
be applied to our sample as the two radio-loud QSOs in the sample
each feature only a compact flat-spectrum radio source.
Also, the size of the NLR has been proposed as a 
straightforward geometric measure
of AGN lifetimes \citep{mar04}.
However, the NLR sizes for the five QSOs of our
sample have not yet been determined.

Thus, as we do not have a direct way to estimate the time elapsed since 
the triggering of the nuclear activity for the objects in our sample, it is
difficult to work out the physical relationship between the mergers we detect 
and the fueling of the QSOs.  An overly simplistic scenario, in which each
QSO was triggered by the merger at the time of the starburst episode would
imply QSO lifetimes on the order of a Gyr, in stark contrast to the
theoretical estimates cited above.  Instead, our observations may provide
evidence in support of significant time
delays between tidal interactions and the fueling of the
central black holes, as predicted by hydrodynamic simulations
(see, e.g., \citealt{bar98,spr05b,hop07b}).

However, while QSOs often reside in bulge-dominated galaxies like
the ones presented in this study, they are also frequently found in
young mergers (e.g., \citealt{urr07,guy06,can01,hut94}). If, as the
accumulating evidence suggests, mergers are indeed essential for 
the triggering of QSOs, then our results imply at least one of two 
scenarios:
Either QSO activity is episodic \citep[e.g.,][]{nor88,hop06} and occurs
over longer timescales than previously speculated, or there is a large range of
values for the time delays between the merger and the onset of the nuclear 
activity.

\acknowledgments
We thank the anonymous referee for the valuable comments
helping to improve the paper.
Support for the {\em HST} program 10421 was provided by NASA through
a grant from the Space Telescope Science Institute, 
which is operated by The Association of Universities for 
Research in Astronomy, Inc., under NASA contract No.\ NAS526555.
Additional support was provided by the National Science Foundation,
under grant AST 0507450. 
B. Jungwiert is also supported by the grant LC06014 of the Czech Ministry
of Education and by Research Plan AV0Z10030501 of the Academy
of Sciences of the Czech Republic.
C. Y. Peng is grateful for support from the HIA Plaskett
fellowship and the STScI Institute/Giacconi fellowship.
This research has made use of the NASA/IPAC Extragalactic Database (NED) 
which is operated by the Jet Propulsion Laboratory, 
California Institute of Technology, 
under contract with the National Aeronautics and Space Administration.

\appendix

\section{NOTES ON INDIVIDUAL OBJECTS}
\label{individual}
Here, we give a more detailed description of
each of the four QSOs.
(Note that MC2\,1635+119 was discussed in detail in Paper I.)
Specifically, we review and discuss published
data, compare them with our findings, and include
a description of neighboring objects.
The redshifts of companion galaxies and other galaxies within
the ACS FOV are taken from NED unless stated
otherwise. The full ACS/WFC images are shown in Figure~\ref{acs},
with labels for the galaxies to which we refer in the following discussion.
Note that we cannot show detailed images for each such galaxy;
instead, we refer the interested reader to the {\em HST} archive.

\subsection{PHL\,909}
Based on older images, the host galaxy of the radio-quiet QSO PHL\,909 
(z=0.172, 1\arcsec~$\simeq$ 2.89\,kpc)
was described as an elliptical galaxy \citep{bah96,bah97,tay96,mcl99,
ham02}.
The age of the stellar population from
off-nuclear optical spectra is not well constrained, 
but the spectra are suggestive of an old age \citep{nol01}.
Note, however, that \citet{hug00} report a significant
increase of flux bluewards of a weak 4000-\AA~break,
but cannot determine whether it is due to scattered
light from the QSO or indicative of a younger stellar
population.

BH mass estimates (making use
of $\sigma$, FWHM$_{\rm H\beta}$ and $L_{\rm 5100}$, 
$R$-band luminosity of the host spheroid)
cover a 
range of 0.17--2.5 $\times$ 10$^9$ $M_{\odot}$ \citep{mcl01, wu02, dun03, wu04}. 
\citet{dun03} predict an Eddington ratio ($L_{\rm nuc}$/$L_{\rm edd}$) of 5.2\%.
\citet{chu06} classify PHL\,909 as Lyman limit system.
They mention a number of faint small objects surrounding
the QSO that could be companions. In a K-band image, the galaxy extends
$\sim$10\arcsec~towards a western companion \citep{dun93}.
The optical spectrum of PHL\,909 shows double-peaked 
Balmer emission lines \citep{str03}.
In the radio, PHL\,909 is an unresolved point source
\citep{kuk98}.

The arc-like features at $\simeq$2\arcsec~from the QSO
mentioned by \citet{bah96} 
correspond most likely to the 
central ring- or disk-like structure ($r \simeq$1.8\arcsec~-- 2.5\arcsec)
seen in our deep {\em HST} ACS image.
However, \citet{bah96} were unable to determine whether the features 
were artifacts due to the much lower S/N of their images.

The neighborhood of PHL\,909 is populated with several
objects, some of which look like dwarf galaxies.
Seven brighter, nearby galaxies make PHL\,909 
appear to reside in a galaxy group. 
However, redshifts are known for only 3 objects, 
of which only one object is at the same redshift as the QSO,
the large spiral galaxy (``h'') to the S
(P.A.~$\simeq$ 183$\degr$) 28\arcsec~away ($z$ = 0.169).
The two other objects are foreground galaxies,
one an elliptical galaxy (``b''; $z$ = 0.102)
to the NE (P.A.~$\simeq$ 28$\degr$) at a distance of 18\arcsec,
and one a spiral (``e''; $z$ = 0.124) to the SE 
(P.A.~$\simeq$ 140$\degr$) 20\arcsec~away.
Two lenticular objects with unknown redshift
lie at 31\arcsec~E (``c''; P.A.~$\simeq$ 96$\degr$)
and 12\arcsec~S (``f''; P.A.~$\simeq$ 173$\degr$) projected
distance of the QSO. 
A spiral galaxy to the SW (``g''; P.A.~$\simeq$ 200$\degr$) 
is 20\arcsec~away.
Another bright object is the galaxy that is apparently interacting
with the QSO (see Section~\ref{morphology}).
In the full FOV of the ACS frame (Fig.~\ref{acs}),
redshifts are known for two other galaxies, both
foreground spirals (``i'', $z$ = 0.104, P.A.~$\simeq$ 29.3$\degr$,
distance $\simeq 1.12\arcmin$ and ``j'', $z$ = 0.082, P.A.~$\simeq$ 76$\degr$,
distance $\simeq 2.34\arcmin$).

\subsection{PKS\,0736+01}
PKS\,0736+01 (z=0.191, 1\arcsec~$\simeq$ 3.14\,kpc) 
was classified as a blazar by \citet{ang80},
as a compact flat-spectrum radio quasar by \citet{wal85},
and as an optically violently variable by \citet{bro89}.
The host galaxy was described as an elliptical by \citet{wri98,mcl99,fal00},
and \citet{ham02}.
Off-nuclear optical spectra have been obtained
and fitted by \citet{hug00} and \citet{nol01},
indicating an age of 12 Gyr.
The estimated BH mass 
(from relations with $\sigma$, FWHM$_{\rm H\beta}$ and $L_{\rm 5100}$, 
$R$-band luminosity of the host spheroid) lies between 
0.14 and 1.3 $\times$ 10$^9$ $M_{\odot}$ \citep{mcl01, wu02, dun03}. 
\citet{dun03} give 3.5\% as an estimate of the Eddington ratio.
The optical {\em HST} WFPC2 image of \citet{mcl99} shows three
companions (two to the NW and one to the SE),
but fails to reveal the low surface-brightness nebulosity,
described as highly disturbed, seen in the K-band \citep{dun93}.
In the near-infrared H-band, the host is resolved but round \citep{kot98}.

Several faint objects surrounding PKS\,0736+01 underlie
the isophotes of the ``spiral''. 
Two of these are background objects
as determined from Keck spectra 
(G. Canalizo \& A. Stockton 2008, in preparation): a lenticular-shaped
galaxy at 5.9\arcsec~to the SE (``a''; Fig.~\ref{acs};
P.A.~$\simeq$ 147$\degr$; $z$ = 0.8500)
and an irregular galaxy at a distance of
4.4\arcsec~to the NW (``b''; P.A.~$\simeq$ 307$\degr$; $z$ = 0.3785).
The latter shows a tidal tail that
is extending toward another object to the NW
(``c''; P.A.~$\simeq$ 297$\degr$; distance $\simeq$ 9\arcsec) of
unknown redshift.

Whether the observed faint spiral-like structure 
in PKS\,0736+01 
indeed indicates the presence of a faint spiral host galaxy,
whether it is formed from debris 
of a merger event as speculated in Section~\ref{mergerage}, 
or whether it is actually a combination of both
(a spiral disturbed by accreted material),
our observations clearly show that the host galaxy
of PKS\,0736+01 cannot be considered to be an undisturbed elliptical galaxy.

\subsection{PG\,0923+201}
The radio-quiet QSO PG\,0923+201 (z=0.19, 1\arcsec~$\simeq$ 3.12\,kpc) 
resides in an elliptical host galaxy \citep{bah97,mcl99,mcl00,ham02,guy06}.
It is not detected as a radio source \citep{kuk98}.
From a noisy optical off-nuclear spectrum,
\citet{nol01} estimate a stellar-population age
of about 12 Gyr.
BH mass estimates (using $\sigma$, FWHM$_{\rm H\beta}$ and $L_{\rm 5100}$, 
$R$-band luminosity of the host spheroid) yield 
$M_{\rm BH}$ $\simeq$ 0.11--2.6 $\times$ 10$^9$ $M_{\odot}$ 
\citep{mcl01, wu02, dun03}. 
\citet{dun03} estimate an
Eddington ratio $L_{\rm nuc}$/$L_{\rm edd}$ of 8.8\%.

\begin{figure}
\epsscale{1}
\plotone{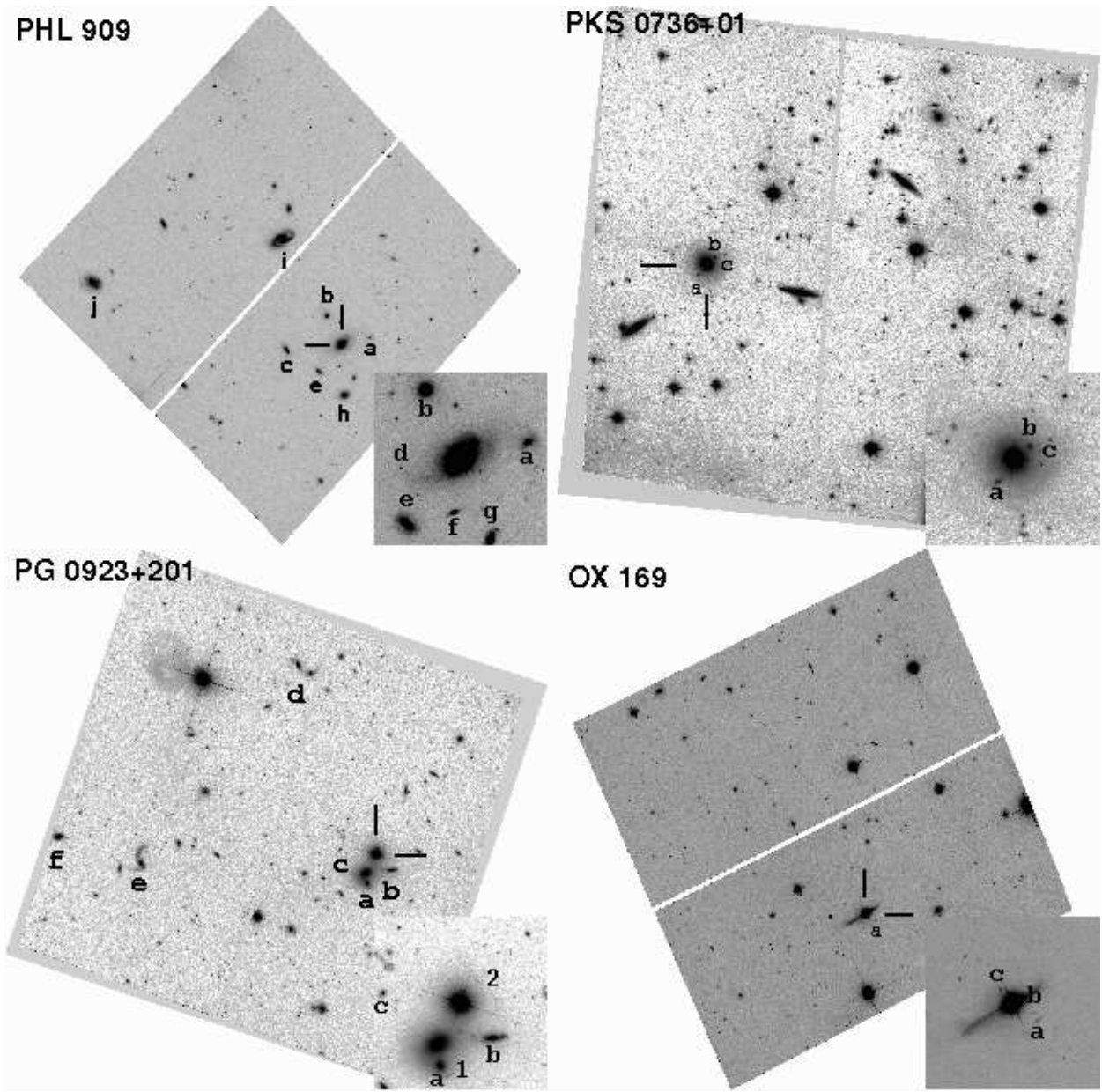}
\caption{Full ACS/WFC images of four QSO host galaxies.
The two bars mark the position of the QSO.
North is up, east is to the left.
The FOV of the ACS/WFC frame is 3.4\arcmin~$\times$ 3.4\arcmin.
As inset, a smaller area around each QSO is shown,
with the image size identical to the images
in Figs.~\ref{final} and~\ref{finalall}.
Letters mark the 
galaxies we refer to in
the text. The letters refer to the closest object north
or south of them.  In the case of PG\,0923+201, the two numbers
refer to the arcs east of them.
}\label{acs}
\end{figure}

PG\,0923+201 is a member of a galaxy group:
In the ACS FOV, at least 5 galaxies 
are at a comparable
redshift of $z \simeq 0.19$,
(pair ``a'', \citet{hec84}; ``c'', \citet{ell91}; pair ``d'',
Keck spectra; G. Canalizo \& A. Stockton 2008, in preparation),
four of which form two interacting pairs (``a'', ``d''; Fig.~\ref{acs}).
The closest pair (``a'') at $\sim$ 12\arcsec~to the east
consists of two large elliptical galaxies
[one galaxy at 10\arcsec~(P.A.~$\simeq$ 153$\degr$) and 
one galaxy at 15\arcsec~~(P.A.~$\simeq$ 161$\degr$) to the SE].
\citet{hut89} speculated that the QSO may be interacting
with these. The northern galaxy of this pair shows a shell-like
structure.
The other pair (``d'') lies at a projected distance of $\simeq 1.6\arcmin$ to
the north-east (P.A.~$\simeq$ 21$\degr$). Both galaxies have
clear tidal tails on one side and are connected by
a tidal bridge.
The fifth galaxy (``c'') at a comparable redshift lies 18\arcsec\ to 
the east (P.A.~$\simeq$ 83$\degr$), a small galaxy with overall elliptical
morphology but faint extended tail-like structures.
The tidal tails make it resemble a more distant and faint
version of the Antennae galaxies
(NGC4038/4039), the best-studied and 
nearest ($\sim$ 19.2 Mpc) example of a major interaction 
between two massive gas-rich spiral galaxies.
However, this object is possibly
at a later stage of the merger as the two nuclei
seem to have coalesced. The SE arm
ends in a knot, a compact object that could be an
interacting dwarf galaxy. 

Note that within the ACS FOV,
another galaxy pair (``e'') at a distance of $\simeq 2\arcmin$~to the east
(P.A.~$\simeq$ 92$\degr$) is at an advanced stage of merging,
with curved tidal tails and the nuclei appearing disrupted in several
fragments. 
While \citet{ell91} list this pair at a redshift of $z = 0.1899$,
Keck spectra show that it is rather at a redshift of $z = 0.2318$
(G. Canalizo \& A. Stockton 2008, in preparation).
The redshift is known for one other galaxy within the ACS FOV, a background spiral
(``f'', $z$ = 0.2328, P.A.~$\simeq$ 87$\degr$,
distance $\simeq 2.6\arcmin$).
There is another very close galaxy to the SW 
(P.A.~$\simeq$ 224$\degr$) of the QSO,
(``b''; distance $\simeq$ 11\arcsec)
with peculiar, tidally disrupted morphology, but unknown redshift.

There are at least two arcs close to the QSO (labeled as ``1'' and ``2''
in Fig.~\ref{acs}) that could be gravitationally lensed galaxies.
However, they may also be tidal debris, especially in the case of ``1'',
which lies close to the interacting pair marked ``a''.

\subsection{OX\,169}
The host galaxy of the radio-loud QSO OX\,169 (z=0.211, 1\arcsec~$\simeq$ 3.39\,kpc) 
has been described as being an elliptical \citep{hut92,tay96,mcl99}.
\citet{nol01} infer an old age for the stellar population of
OX\,169 from their optical off-nuclear spectrum, although
the spectrum has a low S/N ratio.
BH mass estimations using different relations
($\sigma$, FWHM$_{\rm H\beta}$ and $L_{\rm 5100}$, 
$R$-band luminosity of the host spheroid) cover a 
range of 0.22--1.7 $\times$ 10$^9$ $M_{\odot}$ \citep{mcl01, wu02, dun03}. 
The Eddington ratio is estimated to be 6.1\% \citep{dun03}.
From its radio emission, OX\,169 is classified as a compact ($< 4\arcsec$)
flat-spectrum source \citep{fei84}.

OX\,169 seems to have several faint satellite galaxies,
but their redshifts are unknown. 
The closest one at 3.7\arcsec~to the NE
(``c''; P.A.~$\simeq$ 43$\degr$) merges with the outer isophotes of the QSO host galaxy and
resembles a dwarf elliptical galaxy.
Two others lie towards the SW at distances of 
4.6\arcsec~(``b''; P.A.~$\simeq$ 245$\degr$) and
6.1\arcsec~(``a''; P.A.~$\simeq$ 235$\degr$). The former 
has an irregular shape and resembles a tidally disrupted
galaxy, while the latter object is the companion already described in Section~\ref{morphology}.

\onecolumn
\twocolumn

\end{document}